\documentclass[twocolumn,english,aps,prl,showpacs,amssymb,amsfonts]{revtex4}
\usepackage[T1]{fontenc}
\usepackage[latin9]{inputenc}
\usepackage{amsmath}
\usepackage{graphicx}
\usepackage{amssymb}
\usepackage{color}



\usepackage{babel}

\makeatother

\usepackage{babel}
\begin{document}
\def\be{\begin{equation}} 
\def\ee{\end{equation}}
\def\bearr{\begin{eqnarray}}
\def\eearr{\end{eqnarray}}
\def\tc{$T_c~$}
\def\tcx{$\rm T_c(x)~$}
\def\tcl{$T_c^{1*}~$}
\def\lsco{$\rm{La_{2-x}Sr_xCuO_4}$~}
\def\lco{$\rm{La_2CuO_4}$~}
\def\lbco{$\rm{La_{2-x}Ba_x CuO_4}$~}
\def\half{$\frac{1}{2}$~}
\def\dxsq{$\rm d_{x^{2}-y^{2}}$~}
\def\imaxtc{$\rm T_c^{imax}(x)~$}

\title{Superradiant Superconductivity}

\author{G. Baskaran}
\affiliation
{The Institute of Mathematical Sciences, C.I.T. Campus, Chennai 600 113, India\\
Perimeter Institute for Theoretical Physics, Waterloo, Ontario, Canada N2L 2Y5}

\begin{abstract} 
We suggest possibility of Dicke superradiance in superconductors. The necessary 2-level atoms are
identified with  \textit{Anderson pseudo spins} in k-space, seeing a k-dependent self consistent mean field.  A way to couple these 2-level bose atoms to a macroscopically excited coherent boson mode and create  a novel nonequilibrium \textit{superradiant superconductivity} (SRSC) is suggested. Our coherence transfer mechanism offers a hope to realize transient superconductivity, even at room temperatures,  in the pseudo gap phase of certain underdoped cuprates. Recent experiments are briefly discussed in the light of our theory.  Quantum entanglement, QCP and \textit{superfluorescence} properties follow.  
\end{abstract}

\maketitle
{\bf Introduction} Superconductivity is a remarkable macroscopic manifestation of quantum mechanics. A rich physics and phenomenology, including Meissner and Josephson effects are parts of superconductivity \cite{BCS}. Dicke's Superradiance \cite{Dicke1954} is another macroscopic manifestation, exhibited by a collection of 2-level atoms interacting with a single boson mode. The coupled system can develop  quantum coherence, enhanced emission properties and complex dynamics. Certain phenomena in NMR, ESR, optics and cold atoms are related to superradiance. 

In the present work we suggest a way to combine superconductivity and superradiance, We call the resultant  non equilibrium state as \textit{superradiant superconductivity} (SRSC).  In our proposal a macroscopically occupied long wavelength single boson mode interacts with a collection of independent \textit{2-level atoms} located in k-space and creates a Dicke \textit{superradiant} situation, under certain conditions. In this state \textit{certain deformation of Cooper pair wave function is entangled with a coherent external bosonic mode}.
 
Interaction of coherent electromagnetic radiation and ultrasound with superconductor is a 
well studied subject{\cite{MicrowaveExpt,Eliashberg,OwenScalapino,KumarSinha,McIntoshLindesay}. Our proposal of SRSC may have relevance to some known results. An exciting recent development is experimental observation of transient superconductivity well above Tc, induced by certain femtosecond laser pulses, in the pseudo gap phase of cuprates \cite{liscLBCO,liscYBCO}.

In a pioneering theoretical work Eliashberg \cite{Eliashberg} in 1970 showed that microwave induced quasi particle redistribution self consistently enhances gap values and Jc.  Works by Scalapino, Owen and Chang \cite{OwenScalapino}, also focused on quasi particle redistribution.  In a later theory in 1994, McIntosh and Lindesey \cite{McIntoshLindesay} showed that stimulated emission and reabsorption of photon by correlated electron pairs play a fundamental role in superconductivity enhancement. This key insight is one of the triggers for our proposal. Interestingly, in 1968, there was a theoretical suggestion \cite{KumarSinha} for photon induced room temperature superconductivity.

In what follows, we start with an ideal BCS superconductor and show how Dicke superradiance emerges,  when the wavelength of the macroscopically occupied external single boson mode  $\lambda \geq L$, the sample size $L$.  Then we discuss how our mechanism
could gnerate transient superconductivity abouve Tc and discuss recent experiments \cite{liscLBCO} and (see note \cite{liscYBCO} in the light of our mechanism.

In our work we make the tacit assumption that there are suitable relaxation processes involving
quasiparticles and phonons that drains energy to the heat bath efficiently to avoid heating. 
At the same time some energy gets pumped to the electronic sub system to help reach a 
new non equilibrium coherent state for a short time scale. It is the nature of non equilibrium coherent state that we are after. To achieve this we assume that the coherent state of the single boson mode is long lived and does not radiate away  its energy. It exchanges its quanta with the electron subsystem only and gets quantum entangled.  Ours is an \textit{equilibrium statistical mechanics approximation} tailored to get a  glimpse of a remarkable non equilibrium situation. 

{\bf Model}. To develop our theory we follow Anderson's pseudo spin formulation of
BCS theory \cite{PWApseudoSpin}. It helps us to view BCS mean field eigen states as a k-space lattice containing 2-level bose atoms and free fermions. Consider time reversed single particle states $({\bf k}\uparrow,{\bf - k}\downarrow)$, with empty state written as $|0\rangle_{\bf k}$. To generate complete Fock space, we need only 4 states in each
 $({\bf k}\uparrow,{\bf - k}\downarrow)$ : i) $|0\rangle_{\bf k}$, ii) $c^\dagger_{k\uparrow} c^\dagger_{-k\downarrow}|0\rangle_{\bf k}$, iii) $c^\dagger_{k\uparrow} |0\rangle_{\bf k}$ and iv) $c^\dagger_{-k \downarrow} |0\rangle_{\bf k}$. BCS interaction mixes only the 0 and 2-fermion states. Resulting  \textit{ground and excited paired states} are two orthogonal states: $|g\rangle_k \equiv (u_k + v_k c^\dagger_{k\uparrow} c^\dagger_{-k\downarrow}) |0\rangle_k$ and $|e\rangle_k \equiv (u_k c^\dagger_{k\uparrow} c^\dagger_{-k\downarrow}-v_k)|0\rangle_k$. We call these 2-level bosonic states as Anderson atom or \textit{A-atom}. A-atom carries zero total momentum. Single fermion states $c^\dagger_{k\uparrow} |0\rangle_{\bf k}$ and  $c^\dagger_{-k \downarrow} |0\rangle_{\bf k}$,  in   $({\bf k}\uparrow,{\bf - k}\downarrow)$ remain unaffected by BCS interaction.  

An A-atom close to fermi suface is special (see note \cite{note2}). It is a coherent superposition of 0 and 2-electron states. Consequent non zero value of the product $u_k v_k$, around the fermi surface quantifies superconductivity.  

BCS mean field Hamiltonian has the familiar form:
\be
H_{mf} =  \sum \varepsilon_k \alpha^\dagger_{k\sigma} \alpha^{}_{k\sigma},
\ee
where, Bogoliubov quasi particle operators $\alpha^\dagger_{k\sigma} \equiv u_k c^\dagger_{k\sigma} + \sigma v_k c^{}_{-k-\sigma}$ and $\alpha_{k\sigma} \equiv u_k^* c_{k\sigma} + \sigma v_k^* c^{\dagger}_{-k-\sigma}$. The quasi particle energy
$\varepsilon_k \equiv \sqrt {(\frac{\hbar k^2}{2m} - \mu )^2 + \triangle_k^2}$.

Complete set of BCS mean field eigen states can be written as product over all states,
 $({\bf k}\uparrow,{\bf - k}\downarrow)$, each containing either an A-atom in the ground
or excited state or a single upspin or down spin fermion state. Bogoliubov quasi particle operators have very simple action on the BCS eigen states. BCS vacuum, $ |BCS\rangle = \prod_k (u_k + v_k c^\dagger_{k\uparrow} c^\dagger_{-k\downarrow})|0\rangle$ is annihilated by
the annihilation operator, $\alpha_{q\sigma} |BCS\rangle = 0$.

Bogoliubov creation operator, while acting on the BCS ground state, removes an A-atom and 
replaces it by a fermion:
$\alpha^\dagger_{q \uparrow} |BCS\rangle = c^\dagger_{q\uparrow}\prod_{k \neq q}
(u_k + v_k c^\dagger_{k\uparrow} c^\dagger_{-k\downarrow})|0\rangle$ and 
$\alpha^\dagger_{-q \downarrow} |BCS\rangle = c^\dagger_{-q\downarrow}\prod_{k \neq q}
(u_k + v_k c^\dagger_{k\uparrow} c^\dagger_{-k\downarrow})|0\rangle$. 

What is the operator that excites an A-atom ?  Pair of Bogoliubov quasi particle operators 
$\alpha^\dagger_{q \uparrow}\alpha^\dagger_{-q \downarrow}$, 
with total momentum zero and total spin projection zero, acting within $({\bf q}\uparrow,{\bf -q}\downarrow)$ excites an A-atom:
$\alpha^\dagger_{q \uparrow}\alpha^\dagger_{-q \downarrow}|BCS\rangle =
(u_q c^\dagger_{q\uparrow} c^\dagger_{-q\downarrow}-v_q)\prod_{k\neq q}
(u_k + v_k c^\dagger_{k\uparrow} c^\dagger_{-k\downarrow})|0\rangle$. 

The 2-level (bosonic) A-atom subspace can be studied using pseudo spin (Pauli) operators. 
Pseudo spin operators (see note \cite{note3}) are defined as, $\sigma^z_k \equiv
(1 - \alpha^\dagger_{k\uparrow} \alpha^{}_{k\uparrow}
- \alpha^\dagger_{-k\downarrow} \alpha^{}_{-k\downarrow}), ~ \sigma^+_k \equiv
\alpha^\dagger_{k \uparrow}\alpha^\dagger_{-k \downarrow} $ and
$\sigma^-_k \equiv \alpha_{-k \downarrow}\alpha_{k \uparrow}$. 

The BCS mean field Hamiltonian (equation 1) in the boson subspace takes a suggestive form :
\be
H_{mf} = - \sum \varepsilon_k \sigma_k^z ,
\ee
It describes a collection of \textit{non-interacting} pseudo spins in
the prsence of a k-dependent magnetic field of magnitude $\varepsilon_k$. 
Energy level separation
of a 2-level A-atom is 2$\varepsilon_k$.  \textit{Notice that long range interaction in 
k-space in the BCS Hamiltonian leads to free spins in the mean field description, 
but in the presence of a self consistent mean field of magnitude $\varepsilon_k$ in k-space}.
In our pseudo spin basis BCS ground state is a fully aligned ferromagnet, while in Andersons basis pseudo spins twist to form Bloch wall across the fermi surface in k-space (see note \cite{note2}).

Now we consider a simple way to couple A-atoms selectively to a single external boson mode,
with creation and annihilation operators ($b^\dagger, b$).
Interaction of electrons with this mode, in the long wave length (zero momentum transfer)
limit, $\lambda >> L$, where $L$ is the size of the sample, has a simple form:
\be
H_{int} = \frac{1}{{\sqrt N}}\sum B_k (c^\dagger_{k\sigma}c_{k\sigma} + H.c.)(b + b^\dagger)
\ee
Here B$_k$ is a momentum dependent coupling constant and N $\sim$ number of electrons
 in the interaction region.  In terms of Bogoliubov quasiparticle operators, 
\bearr
&~& H_{int}  =\frac{1}{{\sqrt N}} \sum  B_k  (u_k^2 - B_{-k}v_k^2) \alpha^\dagger_{k\sigma} \alpha^{}_{k\sigma}(b + b^\dagger) 
+ ~~~~~\nonumber \\
&+& \frac{1}{{\sqrt N}}\sum (B_k + B_{-k}) u_k v_k (\alpha^\dagger_{k\uparrow}
\alpha^\dagger_{-k\downarrow} + H.c.)(b + b^\dagger)
\eearr
We ignore non resonant terms using rotating wave approximation. Further quasiparticle 
number operators can be also be taken care of using Hartree type of approximations. 
We are left with the important pair annihilation and creation terms:
\be
H_{int} \approx \frac{1}{{\sqrt N}}\sum B_k u_k v_k (\alpha^\dagger_{k\uparrow}
\alpha^\dagger_{-k\downarrow} b_{}  + \alpha^{}_{-k\downarrow}
\alpha^{}_{k\uparrow} b^\dagger)
\ee 
Interms of pseudo spin operators it takes the form $H_{int} \approx
\frac{1}{{\sqrt N}}\sum B_k u_k v_k (\sigma^+_k b_{}  + \sigma^-_k b^\dagger)$. Thus the final form of the
Hamiltonian of the superconductor interacting with a single boson mode is:
\be
H = \hbar \omega_0 (b^\dagger b + \frac{1}{2}) - \sum \varepsilon_k \sigma_k^z + 
\frac{1}{{\sqrt N}}\sum \lambda_k (\sigma_k^+ b + \sigma_k^- b^\dagger)
\ee
where $\lambda_k \equiv (B_k + B_{-k})  u_k v_k$. Equation 5 is a generalized Dicke Hamiltonian \cite{Dicke1954}, where 
\textit{2-level atoms} in k-space have a k-dependent energy level separation, The sum, N$_{t} \equiv$ N$^*$ + N$_{boson}$ of number of excited N$^*$ atoms and number of photons N$_{boson}$, commutes with the Hamiltonian (equation 6)2$\varepsilon_k$.

\textit{Finding a Dicke like Hamiltonian is a key result of our paper}, from which several consequences follow.  

Notice that A-atom-boson mode coupling $\lambda_k \equiv (B_k + B_{-k}) u_k v_k$ is appreciable only in regions where the product  $u_k v_k$ is appreciable. That is, possibility of  superrandiace is intimately connected with pairing phenomenon. The matrix element B$_k$ = - B$_{-k}$for electron-photon coupling. And B$_k$ = + B$_{-k}$ for electron-acoustic phonon coupling \cite{BCS}.  Thus in simple geometries, $\lambda_k = 0$ for electron-electromagnetic radiation coupling.

Our restriction to bosonic subspace and our effective Hamiltonian is a good low temperature  approximation because  i)  k$_B$T << $\Delta_0$, the minimum superconducting gap and density of thermal fermionic quasi particles is small and ii) when $\lambda >> L$, the boson mode  excites only the A-atoms.  

More importantly, we have ignored back reaction, i.e., self consistent modification of u$_k$, v$_k$ or gap function $\Delta_k$, arising from interaction with the boson mode. We will see
later that selfconsistent modification reinforces superradiant superconductivity.

To illustrate superradiance, consider a simple Dicke Hamiltonian, with identical 
two level atoms in resonance with the boson mode,
$H_D = \hbar \omega_0 (b^\dagger b^{} + \frac{1}{2}) - \frac{\hbar \omega_0}{2} \sum_{i} \sigma^z_{i}
 +  \frac{g}{\sqrt{N}} \sum_{i} (b^\dagger \sigma^-_{i} + b^{} \sigma^+_{i})$.
For every value 
of N$_t$ there is an unique ground state, a \textit{nodeless in phase superposition of degenerate 
states with real positive coefficients}. The ground state is a superradiant state capable of 
undergoing a spontaneous emission with an emission strength that scales as N$^2_t$.

For our purpose consider a superconductor at T = 0 in the presence of a macroscopically 
occupied single boson mode $|N_b\rangle$ and allow the coupled system to evolve in time. 
When $\hbar \omega_0$  start increasing towards $\triangle_0$, minimum of the two quasi particle 
gap, a set of k-points which are near resonance with energy of a boson quanta actively participate in superradiance and modify the ground state wave function.  Net density of these active A-atoms depend on quasi particle density of states and the coupling constant $\lambda_k$.

Dicke Hamiltonian, equation 6,  admits Bethe Ansatz solution\cite{BetheAnsatz} for k-independent 
$\lambda_k = \lambda_0$. Using the approximation, $\lambda_k \approx \lambda_0$ for our 
set of near resonant A-atoms, our ground state wave function has Bethe Ansatz form:
\bearr
|SRSC\rangle &\sim & ( b^\dagger + \sum_k w_k ~ \alpha^\dagger_{k\uparrow} \alpha^{\dagger}_{-k\downarrow})^{N_b} 
|BCS\rangle \otimes |0_b\rangle \nonumber \\ 
& \equiv & ( b^\dagger + \sum_k w_k ~ \sigma^+_k )^{N_b} 
|BCS\rangle \otimes |0_b\rangle
\eearr
Here $|0_b\rangle$ is the vacuum of the single boson mode.
Superradiance mixes (hybridizes or entangles) two nearly degenerate neutral modes. One is the single mode
external Bose oscillator. Second is a coherent sum of zero momentum Bogoliubov pair excitations, 
$\sum_k w_k \alpha^\dagger_{k\uparrow} \alpha^{\dagger}_{-k\downarrow}$; or equivalently
an Anderson \textit{pseudo spin wave packet mode} in k-space. It is easy to show that the second boson mode is a dynamic deformation of the Cooper pair wave function (in the relative coordinate of the two electrons, characterized by w$_k$). The center of mass degree of freedom of the Cooper pairs, and hence the phase of superconducting order parameter is not directly influenced by superradiance phenomenon.

Superradiance effect in an s-wave superconductor is maximum, when the boson frequency $\hbar \omega_0$ passes through minimum gap 2$\Delta_o$, where quasi particle density of states has a maximum. If a superconductor supports excited Cooper bound states below 
$2\Delta_0$, depending on the symmetry of the excited states, there will be enhanced superradiance around these bound state energies.

It follows from our work that one should be able to see i) a well known quantum phase transition \cite{HeppLieb}, as a function $\omega, \lambda$ ii) enhanced quantum entanglement \cite{TobiasEntanglement} around the transition point and iii) superfluorescence \cite{Superfluorescence}. 

{\bf Application to Pseudogap Phase of Cuprates} Having theoretically suggested possibility of Dicke superradiance in a BCS superconductor, 
we will address recent experimental observation \cite{liscLBCO,liscYBCO} of femtosecond laser induced transient 
superconductivity in the pseudogap normal state of some cuprates. In the two experiments two different Cu-O bond stretching modes are resonantly excited by an 80 meV  (~$\sim$ 20 THz) femtosecond laser.  In view of resonance, laser pumps its energy and coherence to the infrared phonon mode. Electronic subsystem receives its energy and coherence from the infrared mode. We have a phonon-photon \textit{polariton} Hamiltonian:
\be
H = \hbar \omega_0 a^{\dagger} a + \hbar \omega_0 b^{\dagger} b + 
g (a^{\dagger} b + H.c.) 
\ee
Here $(a^\dagger,a)$ and $(b^\dagger,b)$ are the photon and phonon operators respectively.
As wavelengths of 20 THz infrared radiation and the optic modes is $\sim 150 $ microns, we will approximate the wavelengths by size of the sample. The phonon optical mode coupling `g' 
is of the order of 10 meV. This coupling will lead to interesting Rabi oscillation between two modes, after the femtosecond photon pulse impinges on the superconducting crystal.

It is safe to assume that Cu-O stretching lattice modes in both experiments modulate 
i) site energy and ii) the hopping matrix element `t' of the tight binding electronic 
Hamiltonian for cuprates. To leading order in the normal coordinate displacement u of 
this mode we have $ t = t_0 + \frac{\partial t}{\partial u}|_0 u \equiv t_0 + \alpha_t 
(b^{\dagger} + b) {\rm~~~~ and}$  t$_o$ is the value of hopping integral in the absence of resonant excitation of Cu-O stretching infrared mode. 

As far as the pseudogap normal state of cuprates is concerned, there are experimental evidences \cite{Ong,RamanPseudoGap,STM} and theoretical support \cite{BZAEmeryKivelson} that this metallic state has substantial pairing amplitude and a strong phase fluctuations. It is well described as a 2D vortex liquid above a Kosterlitz-Thouless transition point. This is borne out by Nernst effect \cite{Ong}, Raman effect \cite{RamanPseudoGap}, among other experiments.  In what follows we propose an effective Hamilotonian that is an expression of the fact that  pseudogap phase gap supports local superconductivity. We assumes presence of equal density of positive and negative vortices that are \textit{quasi static and spatially random}. That is, the thermal vortices are slowly moving compared to time scale of interest to us.

Our effective Hamiltonian for pseudogap normal state has the form:
\be
H_{\rm normal} = \sum \varepsilon_m \alpha^\dagger_{\sigma} \alpha^{}_{m\sigma}
\ee
The index m denotes eigen modes of Bogolibov quasiparticle operator
$\alpha^\dagger_{m \sigma} \equiv u_m c^\dagger_{m\sigma} + \sigma v_m c^{}_{-m-\sigma}$
. In view of presence of
disordered vortices in the background,  single particle eigen modes are not Bloch states;
some localized and rest extended. Our conclusions hold good even for the d-wave symmetry situation in cuprates.

In the absence of external magnetic  field we have 
pairs of degenerate single particle eigen states (m$\uparrow$, - m $\downarrow$), connected by time reversal symmetry. As in the BCS case, a pair subspace (m$\uparrow$, - m $\downarrow$) is occupied by A-atom in its ground or excited state or an unpaired fermion. By using same arguments as in the BCS case, bosonic excitation sector in the normal state pseudo gap phase, coupled to the single phonon mode gives us a Dicke type pseudo spin Hamiltonian:
\be
H = \hbar \omega_0 (b^\dagger b + \frac{1}{2}) + \sum \varepsilon_m \sigma^z_m + 
\frac{1}{{\sqrt N}}\sum \lambda_i (\sigma^+_m b + \sigma^-_m b^\dagger)
\ee
Here the operator $\alpha^\dagger_{m\uparrow} \alpha^\dagger_{{ - m}\downarrow} \equiv \sigma^+_m$
excites an A-atom. 

In terms of A-atom and ferminic quasi particle there is a key difference between the BCS supercodnuctor and the cuprate superconductors above Tc. In a standard BCS superconductor, the pair subspace $ ({\bf k}\uparrow,{\bf-k}\downarrow)$, is dominated by  fermionic quasi particles and nearly vanishing density of A-atom.  Whereas, in the pseudogap, which exists over a wide temperature range above Tc,  \textit{the pair subspace (m$\uparrow$, -m $\downarrow$)  is dominated by ground and excited A-atoms and nearly vanishing density of fermions}. This makes pseudogap phase special and susceptible for transient superconductivity.

To understand how superradiance induces transient superconductivity in the pseudo gap phase, we have to go beyond our model Hamiltonian (equation 10) and consider selfconsistent modification of u$_m$ and v$_m$'s.  We offer a \textit{feed back mechanism}. Qualitatively it is as follows.  A subspace (m$\uparrow$, - m $\downarrow$) contains A-atoms with high probability, in ground or excited states; fermions with low probability.  A fraction of excited A-atoms are in resonance with the macroscopically occupied phonon mode. In view of macroscopic occupancy, the boson mode stimulates the near resonant excited A-atom to emit a boson and reach its ground state. In the process we create an excess population of ground state A-atoms. Increase in density of ground state A-atoms means increased superconducting correlation (increase in magnitude of u$_k$v$_k$); consequently an increase in superradiance interaction. Thus there is a positive feedback, which could establishes a transient long range superconducting order. As pseudo gap phase extends to room temperatures in some of
the underdoped cupraets, our mechamism offers a possibility to observe  room temperature transient superconductivity. This is one more incentive for authors of reference  \cite{liscYBCO} to confirm their exciting observations.

To establish superconductivity in the normal state of a Kosterlitz Thouless superconductor, what we need is only a spatial reorganization of random thermal vortices into either  i) a fluid of bound vortex-antivortex pairs as in Kosterlitz-Thouless phase or  ii) an ordered lattice of positive and negative vortices (see note \cite{note5}). The increased pairing correlation from superradiance increases the core energy of the  thermal vortices and a corresponding increase of vortex pair binding energy. Resulting increase in population of paired vortices help create transient superconductivity.

In addition to superconductors, it will be interesting look for superradiant superfluidity in pairing dominatated fermion systems: superfluid He$^3$, cold atoms, heavy nucleii and nuclear matter.

{\bf Acknowledgement} I thank -  N Kumar, K P Sinha, R K Shankar and R Nityananda for 
early discussions on photoinduced superconductivity; P W Anderson and N P Ong 
for an encouraging discussion; N P Ong for bringing to my attention reference \cite{RamanPseudoGap}; 
B. Keimer for an encouraging information \cite{liscYBCO};
DAE, India for a Raja Ramanna Fellowship. This research was supported by Perimeter Institute for Theoretical Physics.

\end{document}